\documentclass[useAMS,usenatbib,usegraphicx]{mn2e}
\usepackage{times}
\usepackage{amsmath}
\usepackage{amssymb}

\input{psfig.sty}

\newcommand{\AIPS}{{$\cal AIPS\/$}}

\def\qir{$q_{\rm IR}$}
\def\lir{$L_{\rm IR}$}

\def\kms{km\,s$^{-1}$}
\def\Td{$T_{\rm d}$}
\def\Md{$M_{\rm d}$}
\def\Tb{$T_{\rm b}$}
\def\K{{\sc k}}
\def\COtw{$^{12}$CO}
\def\jonezero{$J\!=\!1\!-\!0$}

\def\jthreetwo{$J\!=\!3\!-\!2$}
\def\jfourthree{$J\!=\!4\!-\!3$}

\def\gs{\mathrel{\raise0.35ex\hbox{$\scriptstyle >$}\kern-0.6em
\lower0.40ex\hbox{{$\scriptstyle \sim$}}}}
\def\ls{\mathrel{\raise0.35ex\hbox{$\scriptstyle <$}\kern-0.6em
\lower0.40ex\hbox{{$\scriptstyle \sim$}}}}
\def\m@th{\mathsurround=0pt }
\def\eqalign#1{\null\,\vcenter{\openup1\jot \m@th
 \ialign{\strut\hfil$\displaystyle{##}$&$\displaystyle{{}##}$\hfil
 \crcr#1\crcr}}\,}

\title[Mergers and feedback amongst starbursting radio galaxies]
        {Gas-rich mergers and feedback are ubiqitous amongst starbursting
          radio galaxies, as revealed by JVLA, IRAM PdBI and {\it Herschel}}

\author[Ivison et al.]  {R.\,J.\ Ivison,$^{\! 1,2}$
            Ian Smail,$^{\! 3}$
            A.\ Amblard,$^{\! 4}$
            V.\ Arumugam,$^{\! 2}$
            C.\ De Breuck,$^{\! 5}$
            B.\,H.\,C.\ Emonts,$^{\! 6}$ \and
            I.\ Feain,$^{\! 6}$
            T.\,R.\ Greve,$^{\! 7}$
            M.\ Haas,$^{\! 8}$
            E.\ Ibar,$^{\! 1}$
            M.\,J.\ Jarvis,$^{\! 9}$
            A.\ Kov\'aks,$^{\! 10,11}$
            M.\,D.\ Lehnert,$^{\! 12}$ \and
            N.\,P.\,H.\ Nesvadba,$^{\! 13}$
            H.\,J.\,A.\ R\"ottgering,$^{\! 14}$
            N.\ Seymour$^{6}$ and 
            D.\ Wylezalek$^{5}$
       \vspace*{1mm}\\
       $^1$ UK Astronomy Technology Centre, Science and Technology
            Facilities Council, Royal Observatory, Blackford Hill, Edinburgh EH9 3HJ\\
       $^2$ Institute for Astronomy, University of Edinburgh,
            Blackford Hill, Edinburgh EH9 3HJ\\
       $^3$ Institute for Computational Cosmology, Durham University,
            South Road, Durham DH1 3LE\\
       $^4$ NASA, Ames Research Center, Moffett Field, CA\,94035, USA\\
       $^5$ European Southern Observatory, Karl Schwarzschild
            Stra{\ss}e 2, 85748, Garching, Germany\\
       $^6$ CSIRO Astronomy and Space Science, Australia Telescope
            National Facility, P.O.\ Box 76, Epping, NSW 1710, Australia\\
       $^7$ Department of Physics and Astronomy, University College
            London, Gower Street, London WC1E 6BT\\
       $^8$ Astronomisches Institut, Ruhr-Universit\"at Bochum,
            Universit\"atsstra{\ss}e 150, 44801, Bochum, Germany\\
       $^9$ Centre for Astrophysics Research, Science \&
            Technology Research Institute, University of Hertfordshire,
            Hatfield, Herts AL10 9AB\\
       $^{10}$ California Institute of Technology, 301-17, 1200 East
               California Blvd, Pasadena, CA 91125, USA\\
       $^{11}$ University of Minnesota, Institute for Astrophysics,
               116 Church St SE, Minneapolis, MN 55455\\
       $^{12}$ GEPI, Observatoire de Paris, UMR 8111, CNRS,
               Universit\'{e} Paris Diderot, 5 place Jules Janssen,
               92190, Meudon, France\\
       $^{13}$ Institut d'Astrophysique Spatiale, Universit\'{e} Paris
               Sud 11, Orsay, France\\
       $^{14}$ Leiden Observatory, University of Leiden, P.O.\ Box
               9513, 2300 RA, Leiden, The Netherlands}

\date{Accepted 2012 June 17. Received 2012 June 15; in original form
  2012 May 10}

\pagerange{000--000}

\begin{document}

\maketitle

\begin{abstract}
  We report new, sensitive observations of two $z$\,$\sim$\,3--3.5
  far-infrared-luminous radio galaxies, 6C\,1909+72 and
  B3\,J2330+3927, in the $^{12}$CO \jonezero\ transition with the Karl
  Jansky Very Large Array and at 100--500\,$\mu$m using {\it
    Herschel}, alongside new and archival \COtw\ \jfourthree\
  observations from the Plateau de Bure Interferometer.  We introduce
  a new colour-colour diagnostic plot to constrain the redshifts of
  several distant, dusty galaxies in our target fields. A bright SMG
  near 6C\,1909+72 likely shares the same node or filament as the
  signpost active galactic nuclei (AGN), but it is not detected in
  \COtw\ despite $\sim$20,000\,\kms\ of velocity coverage. Also in the
  6C\,1909+72 field, a large, red dust feature spanning
  $\approx$500\,kpc is aligned with the radio jet. We suggest several
  processes by which metal-rich material may have been transported,
  favouring a collimated outflow reminiscent of the jet-oriented metal
  enrichment seen in local cluster environments. Our interferometric
  imaging reveals a gas-rich companion to B3\,J2330+3927; indeed, all
  bar one of the eight $z\gs 2$ radio galaxies (or companions)
  detected in \COtw\ provide some evidence that starburst activity in
  radio-loud AGN at high redshift is driven by the interaction of two
  or more gas-rich systems in which a significant mass of stars has
  already formed, rather than via steady accretion of cold gas from
  the cosmic web. We find that the \COtw\ brightness temperature
  ratios in radio-loud AGN host galaxies are significantly higher than
  those seen in similarly intense starbursts where AGN activity is
  less pronounced. Our most extreme example, where $L'_{\rm
    CO4-3}/L'_{\rm CO1-0}>2.7$, provides evidence that significant
  energy is being deposited rapidly into the molecular gas via X-rays
  and/or mechanical (`quasar-mode') feedback from the AGN, leading to
  a high degree of turbulence {\it globally} and a low optical depth
  in \COtw\ -- feedback that may lead to the cessation of star
  formation on a timescale commensurate with that of the jet activity,
  $\ls$10\,Myr.
\end{abstract}


\section{Introduction}
\label{intro}

High-redshift radio galaxies (HzRGs) are typically identified via
their ultra-steep-spectrum radio emission ($\alpha<-1$ where
$S_{\nu}\propto \nu^{\alpha}$) in flux-limited surveys
\citep*[e.g.][]{tielens79, rottgering94}. Despite the inevitable and
extreme youth of the radio jets that draw our attention to these
distant galaxies, they are associated with the most massive stellar
populations of any known galaxy class -- and presumably the most
massive black holes and host galaxies -- out to the highest redshifts
\citep*{best98, br99, seymour07}.

In the submillimetre (submm; rest-frame far-infrared, FIR) regime,
HzRGs were first explored using relatively primitive submm detectors
such as UKT14 on the James Clerk Maxwell Telescope
\citep[e.g.][]{dunlop94}. Later, with the advent of sensitive submm
cameras, their submm luminosities (and hence dust masses) were found
to be a strongly increasing function of redshift \citep{archibald01,
  reuland04}, even beyond the peak epoch of activity for submm
galaxies \citep[SMGs, $z\sim\rm 1$--3 --][]{chapman05}.

As rare, massive systems, HzRGs are often employed as signposts to
what are expected -- based on our understanding of how cosmic
structures form and evolve \citep[e.g.][]{davis85} -- to be over-dense
regions of the early Universe. Surveys found excesses of various
galaxy types around HzRGs, including Lyman-break galaxies,
Ly\,$\alpha$ emitters and SMGs \citep[][]{ivison00, miley04, greve07,
  venemans07, overzier08, hatch11a}. \citet{stevens03} presented submm
imaging of seven HzRGs, several of which appear to contain extended
($\sim$5--20\,arcsec, $\sim$35--150\,kpc) dust emission, co-spatial
with similarly extended UV emission in several cases \citep{hm93,
  hatch08}. This suggests that obscured starbursts in these over-dense
regions at $z\gs 3$ may differ from the compact ($\ls$0.5\,arcsec, or
$\ls$4\,kpc) events seen in local ultraluminous IR galaxies (ULIRGs)
and, indeed, as seen by high-resolution \COtw\ and radio continuum
imaging of the general $z\sim\rm 2$ SMG field population
\citep{tacconi08, bi08, younger08}. From this we might conclude that
the mechanism for the formation of the very massive galaxies in these
over-dense regions may be fundamentally different to that of the
$\gs$L$_{\star}$ galaxies forming from SMGs \citep{smail04}, although
the earliest \COtw\ \jonezero\ imaging of SMGs
\citep[e.g.][]{carilli11, ivison11, riechers11a, riechers11b, hodge11}
using the Karl Jansky Very Large Array (JVLA) \citep{perley11}
suggests their cold gas may be more extended than the high-$J$ \COtw\
tracers employed previously.

Valiant, early attempts to detect \COtw\ towards HzRGs with
single-dish telescopes ended in failure \citep[e.g.][]{evans96,
  vanojik97}, but the stable spectral baselines afforded by
interferometers eventually allowed the secure detection of \COtw\
\jthreetwo\ towards 53W002 at $z=2.39$ \citep*{scoville97, alloin00}
and \COtw\ \jfourthree\ towards 4C\,60.07 at $z=3.79$ and 6C\,1909+72
at $z=3.53$ \citep[][hereafter P00]{papadopoulos00hzrg},
\defcitealias{papadopoulos00hzrg}{P00} B3\,2330+3927 at $z=3.09$
\citep[][hereafter DB03]{debreuck03}, \defcitealias{debreuck03}{DB03}
TN\,J0121+1320 at $z=3.52$ \citep*{debreuck03ar} and 4C\,41.17 at
$z=3.80$ \citep{debreuck05}.  Detections of the \COtw\ \jonezero\ line
-- that most sensitive to cold gas \citep{pi02} -- are particularly
difficult towards radio galaxies: synchrotron emission often dominates
at the frequency of the line, $\sim$115\,GHz, which is expected to be
roughly an order of magnitude fainter than \COtw\ \jfourthree. The
robust detection of \COtw\ \jonezero\ by \citet{emonts11} towards
MRC\,0152$-$209 is the exception that proves the rule, albeit at a
modest redshift ($z=1.92$). More often, \jonezero\ has proved elusive
\citep{ivison96, papadopoulos05} or detections have been of a
tentative nature \citep*[e.g.][]{greve04, klamer05}.

Here, we present new observations of two HzRGs, 6C\,1909+72 (also
known as 4C\,72.26 and TXS\,J1908+7220) and B3\,J2330+3927
(\citealt{pentericci00}, \citetalias{debreuck03}) using JVLA, the
Institut de Radioastronomie Millim\'{e}trique's Plateau de Bure
Interferometer (IRAM PdBI) and {\it Herschel}\footnote{{\it Herschel}
  is an ESA space observatory with science instruments provided by
  European-led Principal Investigator consortia and with important
  participation from NASA.}  \citep{pilbratt10}.  The PACS and SPIRE
instruments \citep{griffin10, poglitsch10} aboard {\it Herschel} offer
unprecedented sensitivity, with resolution well-matched to
ground-based predecessors such as SCUBA \citep{holland99}. They cover
the decade of wavelengths from 70 to 500\,$\mu$m with
$\lambda/\Delta\lambda\sim 3$, which allows us to probe the peak of
the spectral energy distribution (SED) of a dusty galaxy out to $z\sim
4$, thereby determining its star-formation rate \citep[SFR, via its
FIR luminosity, $L_{\rm IR}$, across rest-frame 8--1,000\,$\mu$m,
e.g.][]{kennicutt98a} and its characteristic dust temperature,
\Td. Beyond the local Universe, these quantities have only rarely been
well constrained until now. Alternatively, we can adopt a reasonable
value for \Td\ and then estimate the redshifts of the SMGs seen in the
vicinity of HzRG signposts \citep[e.g.][]{stevens03} using their
rest-frame FIR colours \citep[e.g.][]{eales03, greve08, penner11},
i.e.\ without resorting to conventional spectroscopy, thereby
determining the likelihood that they inhabit a massive structure
alongside the HzRG.

In the next section we describe an extensive set of observations, then
present the reduced images, spectra and associated analysis in
\S\ref{results}. \S\ref{discussion} contains our interpretation of
those data and discussion of their implications. We finish with our
conclusions in \S\ref{conclusions}. Throughout the paper we use a
cosmology with $H_0 = 71$\,km\,s$^{-1}$\,Mpc$^{-1}$, $\Omega_m=0.27$
and $\Omega_\Lambda = 0.73$.

%
%
\begin{table*}
  \centering
  \caption{\COtw\ emission-line properties.\label{tab:coprop}}
  \begin{tabular}{lcccccc}
    \hline \hline
Target &R.A.&Dec.            &Peak $S_{\nu}$&$z$&{\sc fwhm}&$I_{\rm CO}$\\
&(J2000)&(J2000)&(mJy)            &      &(\kms)&(Jy\,km\,s$^{-1}$)\\
\hline
\jonezero:&&&\\
6C\,1909+72&19:08:23.70$\pm$0.07&+72:20:11.63$\pm$0.36&$0.26\pm 0.04$&$3.5324\pm 0.0007$&$570\pm 90$&$0.222\pm 0.049$\\
B3\,J2330+3927 (AGN host)&--&--&$3\sigma<0.28$&--&--&$3\sigma<0.074$\\
JVLA\,J233024.69+392708.6 (c)&23:30:24.69$\pm$0.02&+39:27:08.58$\pm$0.16&$0.21\pm 0.04$&$3.0884\pm 0.0010$&$720\pm 170$&$0.162\pm 0.034$\\
\hline
\jfourthree:&&&\\
6C\,1909+72&19:08:23.73$\pm$0.02&+72:20:11.54$\pm$0.11&$3.3\pm 0.3$&$3.5324\pm 0.0006$&$800\pm 90$&$2.69\pm 0.27$\\
B3\,J2330+3927 (AGN host)&23:30:24.85$\pm$0.02&+39:27:12.04$\pm$0.15&$2.7\pm 0.3$&$3.0934\pm 0.0005$&$830\pm 100$&$3.29\pm 0.51$\\
JVLA\,J233024.69+392708.6 (c)&23:30:24.62$\pm$0.02&+39:27:08.46$\pm$0.18&$1.9\pm 0.4$&$3.0901\pm 0.0006$&$520\pm 110$&$1.12\pm 0.17$\\
\hline
 \end{tabular}
\end{table*} 

\section{Observations}
\label{observations}

The observations described hereafter targeted the HzRGs, 6C\,1909+72
and B3\,J2330+3927. The former presents a classic double-lobed (plus
core) radio morphology, subtending $\sim$15\,arcsec, and has a total
1.4-GHz flux density, $S_{\rm 1.4GHz}\sim 259$\,mJy; the latter is
brighter, $S_{\rm 1.4GHz}\sim 405$\,mJy, with what appears at first
sight to be a similar morphology, if more compact ($\sim$2\,arcsec),
but which high-resolution radio imaging revealed to be an unusually
one-sided jet driven by a compact, flat-spectrum core
\citep{perez-torres05}.  Alongside PKS\,1138$-$262, 6C\,1909+72 and
B3\,J2330+3927 are the most luminous of the 69 HzRGs explored by
\citet{seymour07} in the rest-frame near-IR using {\it Spitzer}.

\subsection{JVLA observations}
\label{jvlaobs}

Whilst the JVLA was in its most compact configuration during 2011
September--November, we acquired amongst the first data taken in a new
mode offering almost an order of magnitude more bandwidth than was
previously possible, tuning to the \COtw\ \jonezero\ transition at
115.27120256\,GHz \citep{mn94} for our target HzRGs.

Short slots, usually 2--3\,hr long, were scheduled dynamically to
ensure excellent phase stability and transparency in the K and Ka
atmospheric windows for 6C\,1909+72 and B3\,J2330+392,
respectively. During these slots, data were recorded\footnote{A bug in
  the Correlator Back End computer during this period resulted in only
  1\,sec of data being saved per integration, meaning that a
  significant fraction of the useful data were lost.}  every 3\,sec in
each of $2\times 8$ contiguous baseband (dual-polarisation) pairs,
each baseband comprising $64\times 2$-MHz channels for a total
dual-polarisation bandwidth of 2,048\,MHz, well over 20,000\,\kms\ at
the redshift of 6C\,1909+72. Bright, compact calibration sources --
flat-spectrum blazars lying within a few degrees of our target
galaxies -- were observed every few minutes to determine accurate
complex gain solutions and bandpass corrections. 3C\,48 and 3C\,286
were also observed to set the absolute flux scale, and the pointing
accuracy was checked locally every hour.

For 6C\,1909+72 the two sets of eight contiguous baseband pairs were
themselves placed contiguously. We shifted the \COtw\ \jonezero\ line
(expected at 25.434\,GHz for $z=3.5322$) by 180\,MHz from the centre
of the available bandwidth to avoid the edge of a baseband, or one of
the end basebands.

For B3\,J2330+3927, the \COtw\ \jonezero\ line lies below 32\,GHz in the
Ka band. As such, only one set of eight contiguous baseband pairs
could be deployed on the line (expected at 28.156\,GHz for $z=3.094$,
but purposefully offset by 64MHz to avoid the edge of a baseband); the
other set of eight basebands were tuned to 32.5\,GHz.

The data were reduced and imaged using \AIPS\ following the recipes
described by \citet{ivison11}, though with a number of significant
changes: data were loaded using {\sc bdf2aips}, avoiding any
compression, and {\sc fring} was used to optimise the delays, in
software, based on 1\,min of data for 3C\,48 or 3C\,286. The basebands
were knitted together using the {\sc noifs} task, yielding $uv$
datasets with either 1,024 or 512 $\times$ 2-MHz channels. Continuum
images were made, using all of these data to obtain very deep maps of
the synchrotron emission at 25.4 and 28.2\,GHz (for 6C\,1909+72 and
B3\,J2330+3927, respectively), which was then cleaned and subtracted
from the $uv$ data (using {\sc uvsub}). The channels were then imaged
in groups of four (yielding $\sim$90\,km\,s$^{-1}$ velocity
resolution) over $\sim$3-arcmin-diameter fields, with natural
weighting ({\sc robust = 5}), to form large cubes centred on the radio
galaxies. The resulting spatial resolutions were $4.0\times
3.0$\,arcsec$^2$ at position angle (PA), 171$^{\circ}$, for
6C\,1909+72 and $3.4\times 2.8$\,arcsec$^2$ at PA, 72$^{\circ}$, for
B3\,J2330+3927.

\subsection{IRAM PdBI imaging}
\label{pdbiobs}

Observations of 6C\,1909+72 were made in \COtw\ \jfourthree\ using
PdBI's C and D configurations in the 3-mm atmospheric window during
2007 July (with five antennas) and December (with six antennas) in a
series of $\sim$6--11-hr tracks with good phase stability
(0.6--1.0\,arcsec seeing) and transparency (1.5--2.5\,mm of water
vapour). The local calibrator, 1928+738, was used to track amplitude
and phase, with 1642+690 used for an independent check. Absolute
fluxes were calibrated using MWC\,349. Data cubes were made via \AIPS\
using a natural weighting scheme, yielding a $3.1\times
2.7$\,arcsec$^2$ beam with the major axis at a PA of 47$^{\circ}$.

The original PdBI data acquired by \citetalias{debreuck03} for
B3\,J2330+3927 were re-imaged using \AIPS\ using a natural weighting
scheme, discarding those channels observed only in the compact
configuration, yielding a $3.2\times 2.4$\,arcsec$^2$ beam with the
major axis at a PA of 72$^{\circ}$.

\subsection{{\it Herschel} imaging}
\label{herschelobs}

The {\it Herschel} data presented in this paper comprise a small part
of the {\it Herschel} Radio Galaxy Evolution Project (HeRGE) OT1
programme \citep{seymour12}. With PACS, we made near-orthogonal scans
of our HzRG targets at nominal speed (i.e.\ with the telescope
tracking at 20\,arcsec\,s$^{-1}$) using mini-map mode, spending a
total of 288\,sec on source, recording data simultaneously at 100 and
160\,$\mu$m, each scan comprising $12 \times 4$-arcmin scan legs
separated by 4\,arcsec. With SPIRE we approached the confusion limit
\citep{nguyen10} with just three repetitions (111\,sec on source) of
its `small map' mode, obtaining data simultaneously at 250, 350 and
500\,$\mu$m. The SPIRE images were produced using the standard
pipeline; the PACS data were tackled with a variant of the pipeline
developed by \citet{ibar10pacs}.

Flux densities for SPIRE were measured using {\sc sussextractor}
\citep{smith12}, as implemented in the {\it Herschel} Data Processing
System. For PACS, we used apertures of radius 9 and 13.8\,arcsec at
100 and 160\,$\mu$m, respectively, with appropriate aperture
corrections, with errors determined by placing many random apertures
in regions of the image with integration times similar to those of our
targets.

%
%
\begin{figure*}
\centerline{\psfig{file=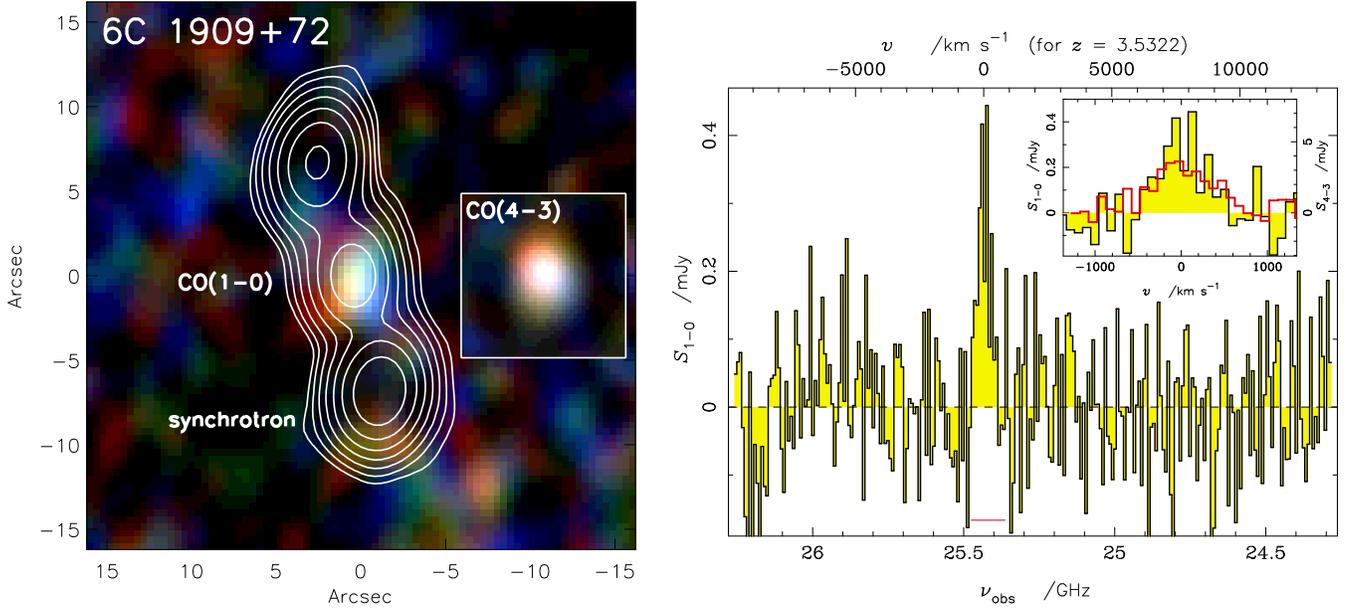,width=3.3in,angle=0}
\hspace*{3mm} 
\psfig{file=figs/1909-spectrum.eps,width=3.5in,angle=270}}
\vspace*{-5mm} 
\noindent{\small\addtolength{\baselineskip}{-3pt}} 
\caption{{\it Left:} False colour image of \COtw\ \jonezero\ emission
  from 6C\,1909+72, as measured by JVLA. N is up; E is left. Inset:
  \COtw\ \jfourthree\ as measured at IRAM PdBI, on the same spatial
  scale. Emission coming towards and going away from us relative to
  the systemic velocity is represented with the appropriate colours,
  revealing hints of a velocity field common to both \COtw\ lines,
  though not globally. The much brighter 25-GHz synchrotron emission,
  against which we have struggled to discern the \COtw\ \jonezero\
  emission, is represented with isophotal contours at $-$3, 3, 6,
  12... $\times$ the local noise level. {\it Right:} \COtw\ \jonezero\
  spectrum of 6C\,1909+72, after subtraction of the dominant
  synchrotron emission component, smoothed with a 140-\kms\ {\sc fwhm}
  Gaussian. Inset: zoomed in on the line, with the new IRAM PdBI
  \COtw\ \jfourthree\ spectrum (\S\ref{pdbiobs}) shown in red, binned
  to 100\,\kms\ and scaled by $16^{-1}\times$ to be on the same
  Rayleigh-Jeans brightness temperature (\Tb) scale as the \COtw\
  \jonezero\ spectrum. The velocity scales correspond to $z=3.5322$,
  approximately midway between the two components ($z=3.5203$ and
  $3.5401$) identified by \citet{smith10}. The red horizontal line
  shows the spectral region summed to create the \COtw\ \jonezero\
  image shown alongside, and to determine $I_{\rm CO}$.}
\label{fig:1909}
\end{figure*}

%
%
\begin{figure*}
\centerline{\psfig{file=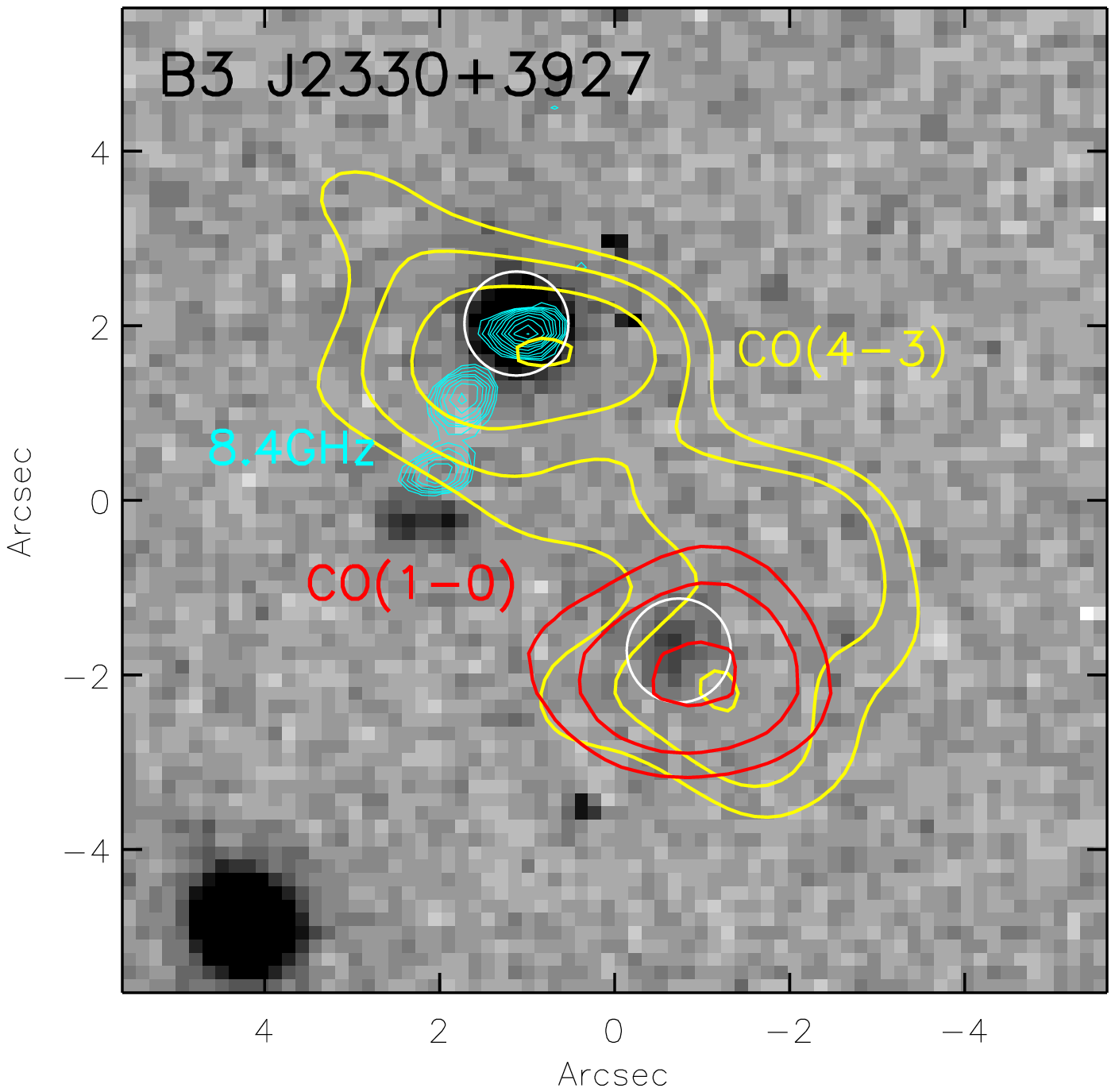,width=3.3in,angle=0}
\hspace*{3mm}
\psfig{file=figs/b3j2330-spectrum.eps,width=3.5in,angle=270}}
\vspace*{-4mm} 
\noindent{\small\addtolength{\baselineskip}{-3pt}} 
\caption{{\it Left:} Greyscale $K'$ image of the B3\,J2330+3927 field
  \citepalias{debreuck03} with isophotal contours of the \COtw\
  \jonezero\ emission in red ($-$3, 3, $3\sqrt{2}$, 6... $\times$
  local noise level). N is up; E is left. The \COtw\ \jonezero\ is
  associated with component c (circled, lower right), rather than the
  core of the AGN host galaxy (also circled, betrayed by its powerful
  synchrotron emission -- cyan contours --
  \citealt{perez-torres05}). \COtw\ \jfourthree\ emission is shown in
  yellow -- a radically different morphology from that presented by
  \citetalias{debreuck03}. The brightest component is centred on the
  AGN while a fainter clump lies to the south-west, coincident with
  component c. The astrometric uncertainties here are
  $\ls$0.5\,arcsec. {\it Right:} \COtw\ \jonezero\ spectrum of
  JVLA\,233024.69+392708.6 (component c), near B3\,J2330+3927. Inset:
  zoomed in on the line, with the IRAM PdBI \COtw\ \jfourthree\
  spectrum of the same galaxy (\S\ref{pdbiobs}) shown in red, binned
  to 53\,\kms\ and scaled by $16^{-1}\times$ to be on the same
  Rayleigh-Jeans \Tb\ scale as the \COtw\ \jonezero\ spectrum. The red
  horizontal line shows the spectral region summed to create the
  \COtw\ \jonezero\ image shown alongside, and to determine $I_{\rm
    CO}$.}
\label{fig:b3}
\end{figure*}

\section{Results}
\label{results}

\subsection{6C\,1909+72}
\label{6c}

%
%
\begin{table*}
  \centering
  \caption{Measured flux densities and properties of HzRGs and brightest field SMGs.\label{tab:fluxes}}
  \begin{tabular}{lcccccc}
    \hline \hline
Target &$S_{\rm 100\mu m}$&$S_{\rm 160\mu m}$&$S^\dagger_{\rm 250\mu
  m}$&$S^\dagger_{\rm 350\mu m}$&$S^\dagger_{\rm 500\mu m}$&$S_{\rm 850\mu
  m}$\\
 & (mJy) & (mJy) & (mJy) & (mJy)& (mJy) & (mJy)\\
   \hline
6C\,1909+72&$17.0\pm3.0$&$34.9\pm6.5$&$57.2\pm2.7$&$69.6\pm2.8$&$63.4\pm3.3$&$34.9\pm3.0$\\
SMM\,J190827.5+721928&$0.0\pm2.9$&$-0.1\pm6.7$&$45.1\pm2.6$&$65.1\pm2.7$&$61.2\pm3.0$&$23.0\pm 2.5$\\
B3\,J2330+3927&$11.3\pm2.9^\ddagger$&$27.0\pm6.2^\ddagger$&$52.0\pm2.7$&$60.3\pm2.7$&$57.3\pm3.3$&$22.2\pm 2.7$\\
SMM\,J233019.1+392703&$5.4\pm3.3$&$4.2\pm11.9$&$31.6\pm2.8$&$35.4\pm2.6$&$21.1\pm3.4$&$8.2\pm 1.9$\\
    \hline
  \end{tabular}

\noindent
$^\dagger$ Uncertainty does not include the contribution from confusion noise,
nor any possible systematic offset in flux calibration.\\
$^\ddagger$ Astrometry is consistent with a significant fraction of the
PACS emission originating from JVLA\,233024.69+392708.6 (Fig.~\ref{fig:b3rgb}).
\end{table*}
 
The synchrotron continuum emission from 6C\,1909+72 dominates the
observed flux density at 25\,GHz, being almost an order of magnitude
brighter than the \COtw\ line peak (Fig.~\ref{fig:1909}, left). To
assess the \COtw\ line properties we subtracted an accurate model of
the continuum emission, exploiting the impressive bandwidth now
available at JVLA.

In the resulting continuum-free JVLA datacube, \COtw\ \jonezero\ line
emission is seen clearly, centred at $z=3.5324\pm 0.0007$
(Fig.~\ref{fig:1909}, right), closely matching that seen in our \COtw\
\jfourthree\ spectrum from IRAM PdBI, $z=3.5324\pm 0.0006$. The basic
characteristics of the \COtw\ line emission -- peak and total flux
densities, line widths, etc.\ -- are listed in Table~\ref{tab:coprop},
for both transitions. It is notable that we measure $I_{\rm
  CO4-3}=2.69\pm 0.27$\,Jy\,\kms, cf.\ the $1.62\pm 0.30$\,Jy\,\kms\
measured by \citetalias{papadopoulos00hzrg}, who may also have
under-estimated the line width due to a lack of available bandwidth.

Both the \COtw\ \jonezero\ and \jfourthree\ emission are partially
resolved, even with a natural weighting scheme. Some velocity
structure can be discerned in Fig.~\ref{fig:1909}, common to both
transitions, but there does not appear to be a strong, coherent
gradient.

From the measured \COtw\ properties, we find $L'_{\rm CO1-0}=(1.19\pm
0.26)\times 10^{11}$\,\K\,\kms\,pc$^2$ and the molecular gas mass,
$M_{\rm gas}$ (H$_2$ + He), is therefore likely around
$\sim10^{11}$\,M$_{\odot}$, naively adopting $M_{\rm gas}/L'_{\rm
  CO}=0.8$\,M$_{\odot}$ (\K\,\kms\,pc$^2$)$^{-1}$ \citep{ds98}.

\subsection{B3\,J2330+3927}
\label{b3}

%
%
\begin{figure}
\centerline{\psfig{file=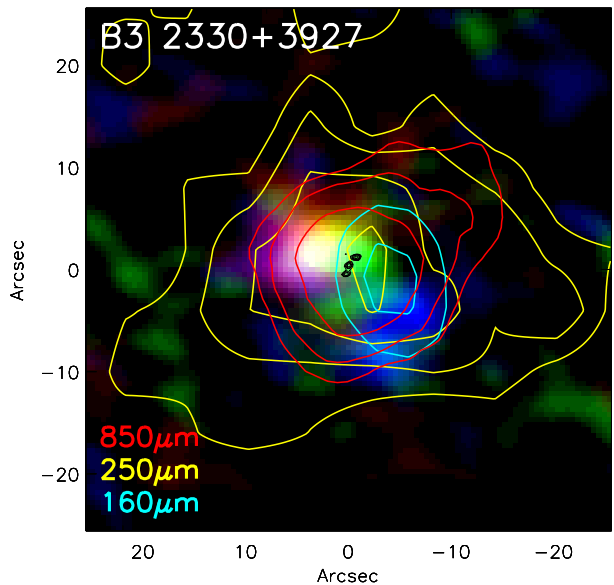,width=3.2in,angle=0}}
\vspace{-4mm} 
\noindent{\small\addtolength{\baselineskip}{-3pt}} 
\caption{Three velocity slices in \COtw\ \jfourthree\ towards
  B3\,J2330+3927 displayed as a false-colour RGB image, showing how
  the super-thermal emission centred near the radio-loud AGN (shown as
  black dots) is offset to the red from the \COtw\ emission associated
  with the neighbouring, gas-rich galaxy,
  JVLA\,233024.69+392708.6. Cyan, yellow and red contours (starting at
  3\,$\sigma$ and spaced by $\sqrt2$) show the 160-, 250- and
  850-$\mu$m continuum emission as seen by {\it Herschel} and SCUBA
  \citep{stevens03}, with no astrometric tweaks applied, consistent
  with cold dust distributed throughout the system. N is up; E is
  left. }
\label{fig:b3rgb}
\end{figure}
 
%
%
\begin{figure}
\centerline{\psfig{file=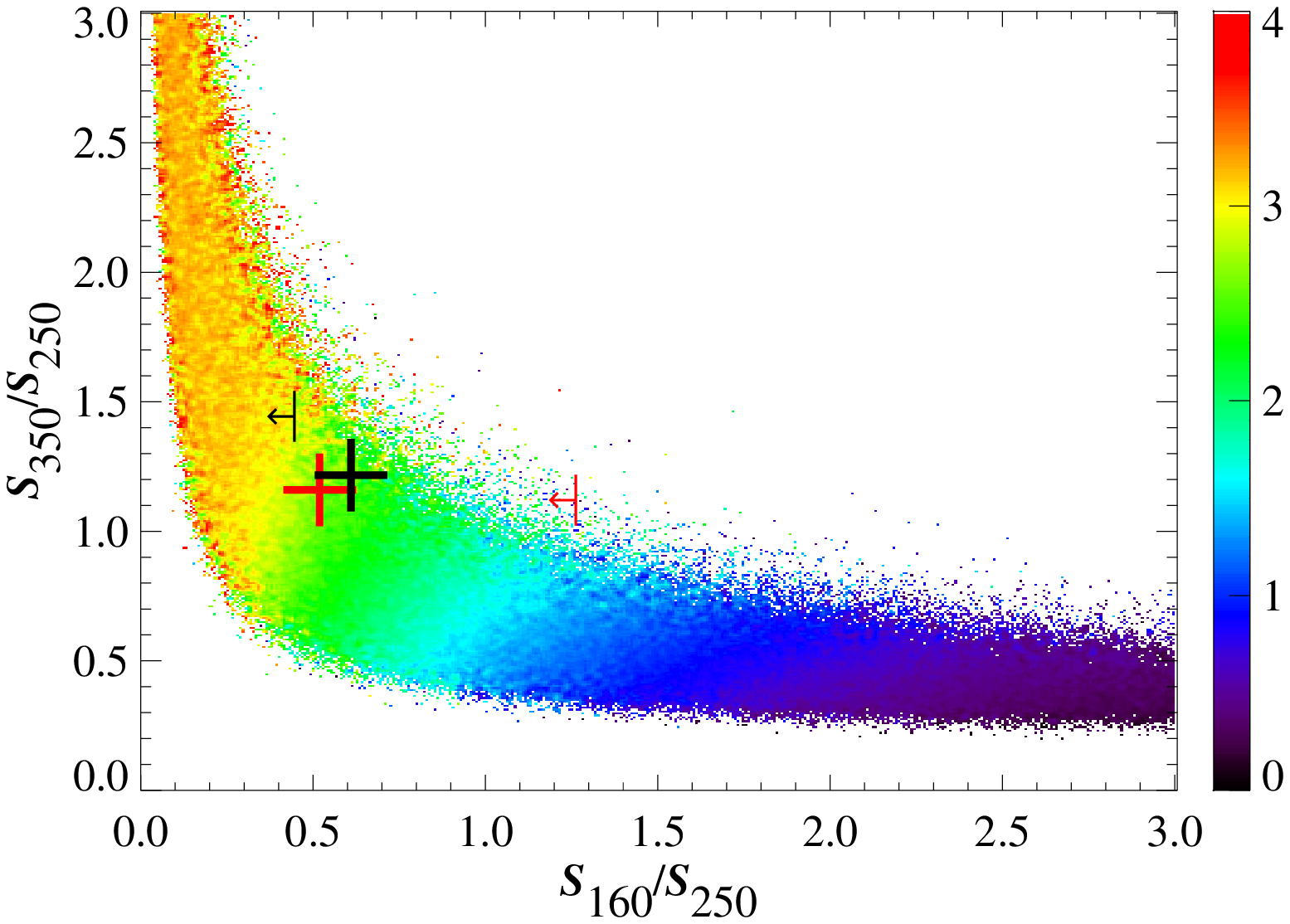,width=3.05in,angle=0}}
\centerline{\psfig{file=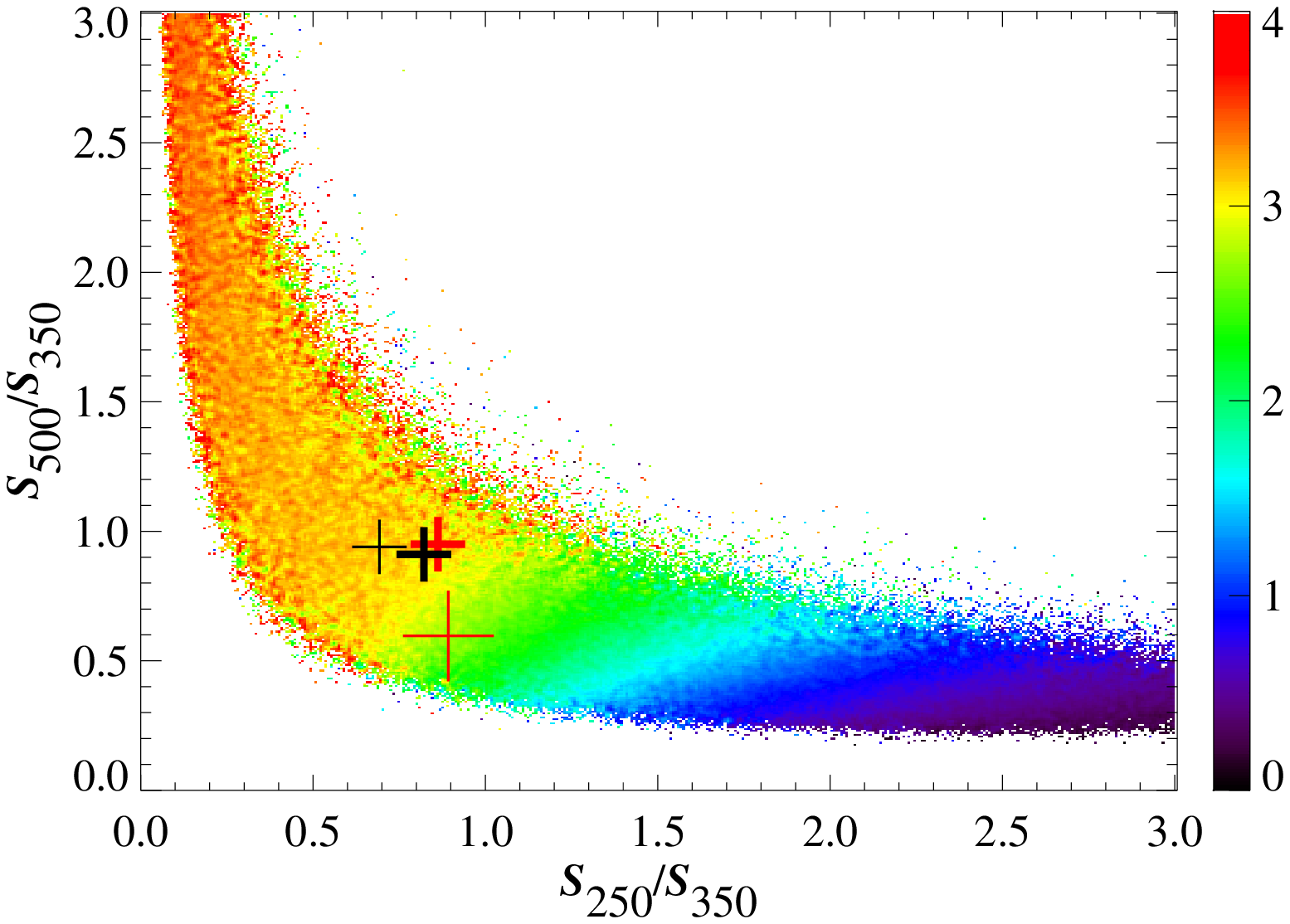,width=3.05in,angle=0}}
\centerline{\psfig{file=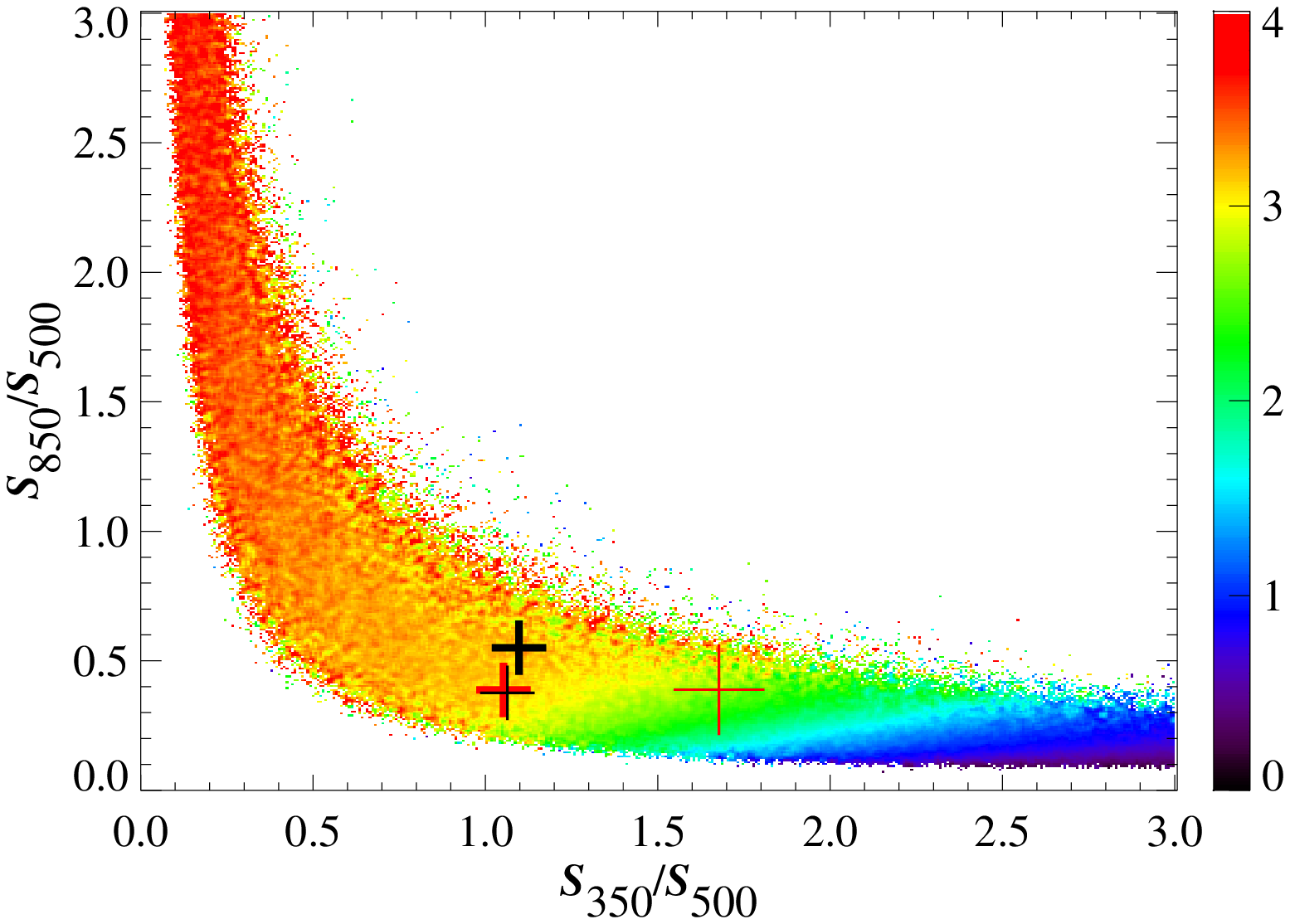,width=3.05in,angle=0}}
\vspace{-5mm} 
\noindent{\small\addtolength{\baselineskip}{-3pt}} 
\caption{Colour-colour plots for the HzRGs and their neighbouring
  SMGs, adapted from \citet{amblard10}, together with a new plot
  designed to exploit our SCUBA 850-$\mu$m photometry. For
  galaxies/bands without robust detections in Table~\ref{tab:fluxes},
  upper limits are shown (calculated as
  $S_\lambda+3\sigma_\lambda$). for The coloured backgrounds indicate
  the redshifts of model SEDs \citep{amblard10}.  The red crosses
  represent B3\,J2330+3927 (thick lines) and SMM\,J233019.1+392703,
  where the latter appears to lie at a lower redshift than the radio
  galaxy. The black crosses represent 6C\,1909+72 (thick lines) and
  SMM\,J190827.5+721928, which have remarkably similar colours,
  consistent with a shared redshift of $\gs$3.}
\label{fig:amblard}
\end{figure} 
 
At the position of the radio galaxy core, as identified by
\citet{perez-torres05} and detected in \COtw\ \jfourthree\ using PdBI
by \citepalias{debreuck03}, we find synchrotron emission ($S_{\rm
  28.2GHz}=5.14\pm0.02$\,mJy centred at $\rm 23^h 30^m 24.^s866,
+39^{\circ} 27' 11.84''$ J2000) but there is no evidence of \COtw\
\jonezero\ emission in a synchrotron-subtracted data cube. The
(3-$\sigma$) limit we can set, assuming the same line width as the
\COtw\ \jfourthree\ emission (830\,\kms\ {\sc fwhm}) and following
Appendix~A of \citet*{sih95} is $I_{\rm CO} < 0.074$\,Jy\,\kms. The
basic characteristics of the \COtw\ line emission are again listed in
Table~\ref{tab:coprop}, for both transitions.

Fig.~\ref{fig:b3} shows the Keck/NIRC $K$-band imaging
\citepalias[from][]{debreuck03} as a greyscale, superimposed with
contours to show the emission in 8.4-GHz continuum (synchrotron),
\COtw\ \jonezero\ and \COtw\ \jfourthree. Approximately 4\,arcsec to
the SSW of the AGN (and $\sim$260\,\kms\ bluewards of its
\COtw\ \jfourthree\ redshift) we do see emission in
\COtw\ \jonezero. It is coincident with what \citetalias{debreuck03}
labelled component `c' and tentatively suggested might lie at the same
redshift as the radio galaxy.  We show its spectra in
\COtw\ \jonezero\ and \jfourthree\ in Fig.~\ref{fig:b3}. Our new
\COtw\ \jonezero\ imaging and our re-analysis of the
\COtw\ \jfourthree\ data make it clear that component c is in fact a
gas-rich galaxy in the immediate vicinity of the radio-loud AGN, which
is presumably interacting with (and thereby triggering activity in)
the host galaxy of the AGN.  Component c is detected in at least three
IRAC bands and so has a significant stellar population. It is not
obvious that it contributes significantly to the Lyman\,$\alpha$ halo
that surrounds B3\,J2330+3927 \citep{matsuda09}.

Fig.~\ref{fig:b3rgb} reveals the dynamical structure of the
\COtw\ emission, with a blue--red gradient leading from component c to
beyond the radio-loud AGN. The continuum emission at $\lambda_{\rm
  obs}=160$, 250 and 850\,$\mu$m (rest-frame 40, 60 and 200\,$\mu$m)
seen by PACS, SPIRE and SCUBA are also shown. At $\lambda_{\rm
  obs}=24\,\mu$m, the emission is predominantly from the AGN; at
$\lambda_{\rm obs}=100$ and 160\,$\mu$m, it is predominantly from
component c; at $\lambda_{\rm obs}\ge 250\,\mu$m the emission is
centred between the two.  We have evidence, then, that dust at a
variety of temperatures is distributed throughout the system, yet we
lack the spatial resolution to disentangle these components reliably.

The total flux in \COtw\ \jonezero\ for component c, $I_{\rm
  CO}=0.162\pm 0.034$\,Jy\,\kms. The excitation situation for
component c is comparable to that of SMGs \citep{harris10, ivison11},
with a brightness temperature (\Tb) ratio, $L'_{\rm CO4-3}/L'_{\rm
  CO1-0} = r_{4-3/1-0} = 0.43\pm 0.11$ \citep[cf.\ $0.41\pm0.07$
--][]{bothwell12}. This differs strikingly from that of the radio
galaxy core, where the \Tb\ ratio is entirely inconsistent with
thermal emission, $r_{4-3/1-0} > 2.7$ ($3\sigma$), i.e.\ the emission
is super-thermal\footnote{Strictly, {\it super-thermal} excitation
  exists only where local thermodynamic equilibrium has been violated
  by a population inversion. Here, we use it as short-hand for $r >
  1$.}  (see \S\ref{super}).

\citetalias{debreuck03} assumed $r_{4-3/1-0}=0.45$ to extrapolate
their \COtw\ \jfourthree\ measurement to \jonezero\ and calculate a
gas mass for B3\,J2330+3927. We can see now that this will have
resulted in an over-estimate of $M_{\rm gas}$ for the AGN host
galaxy. Our limit on $I_{\rm CO1-0}$ implies $L'_{\rm CO} < 2.9\times
10^{10}$\,\K\,\kms\,pc$^2$ and $M_{\rm gas}< 2.3\times
10^{10}$\,M$_{\odot}$ where $M_{\rm gas}/L'_{\rm CO}=0.8$\,M$_{\odot}$
(\K\,\kms\,pc$^2$)$^{-1}$.

For the gas-rich galaxy previously known as component c, which we
re-name JVLA\,233024.69+392708.6, we find $L'_{\rm CO} = (6.9\pm
1.5)\times 10^{10}$\,\K\,\kms\,pc$^2$ and $M_{\rm gas}\sim 5.5\times
10^{10}$\,M$_{\odot}$.

\subsection{On the HzRGs and their environments}
\label{field}

The over-density of bright SMGs found around HzRGs, their faintness at
optical/IR wavebands and the alignment of those SMGs with the jets
from the central AGN were taken as tentative evidence that HzRGs are
signposting massive structures in which other galaxies are also
undergoing intense starbursts \citep{ivison00, stevens03}.

Our deep {\it Herschel} imaging allows us to assess the likely
redshifts of these bright neighbouring SMGs via their FIR/submm
colours \citep[e.g.][though such colours are only sensitive to
$(1+z)/T_{\rm d}$ -- see \citealt{blain99}]{amblard10}. We report
their FIR/submm flux densities in Table~\ref{tab:fluxes}.

The upper panels of Fig.~\ref{fig:amblard} show two of the
colour-colour diagnostic plots employed by \citet{amblard10} to assess
the redshift and \Td\ of galaxies detected by H-ATLAS \citep{eales10},
probing their colours across the rest-frame $\sim$100-$\mu$m bump. The
coloured backgrounds indicate the redshifts of model SEDs. The new,
lower panel of Fig.~\ref{fig:amblard} is based on the same model SEDs
and is designed to exploit information from our SCUBA 850-$\mu$m
imaging that is particularly relevant to galaxies at $z>2$.

The red crosses represent B3\,J2330+3927 and its brightest SMG
neighbour, SMM\,J233019.1+392703. The position of the latter --
particularly in the lower two plots -- suggests strongly that it lies
at a lower redshift than the radio galaxy, nearer the peak of the
radio-detected SMG population at $z\sim 2.2$
\citep[e.g.][]{chapman05}.

%
%
\begin{figure*}
\centerline{\psfig{file=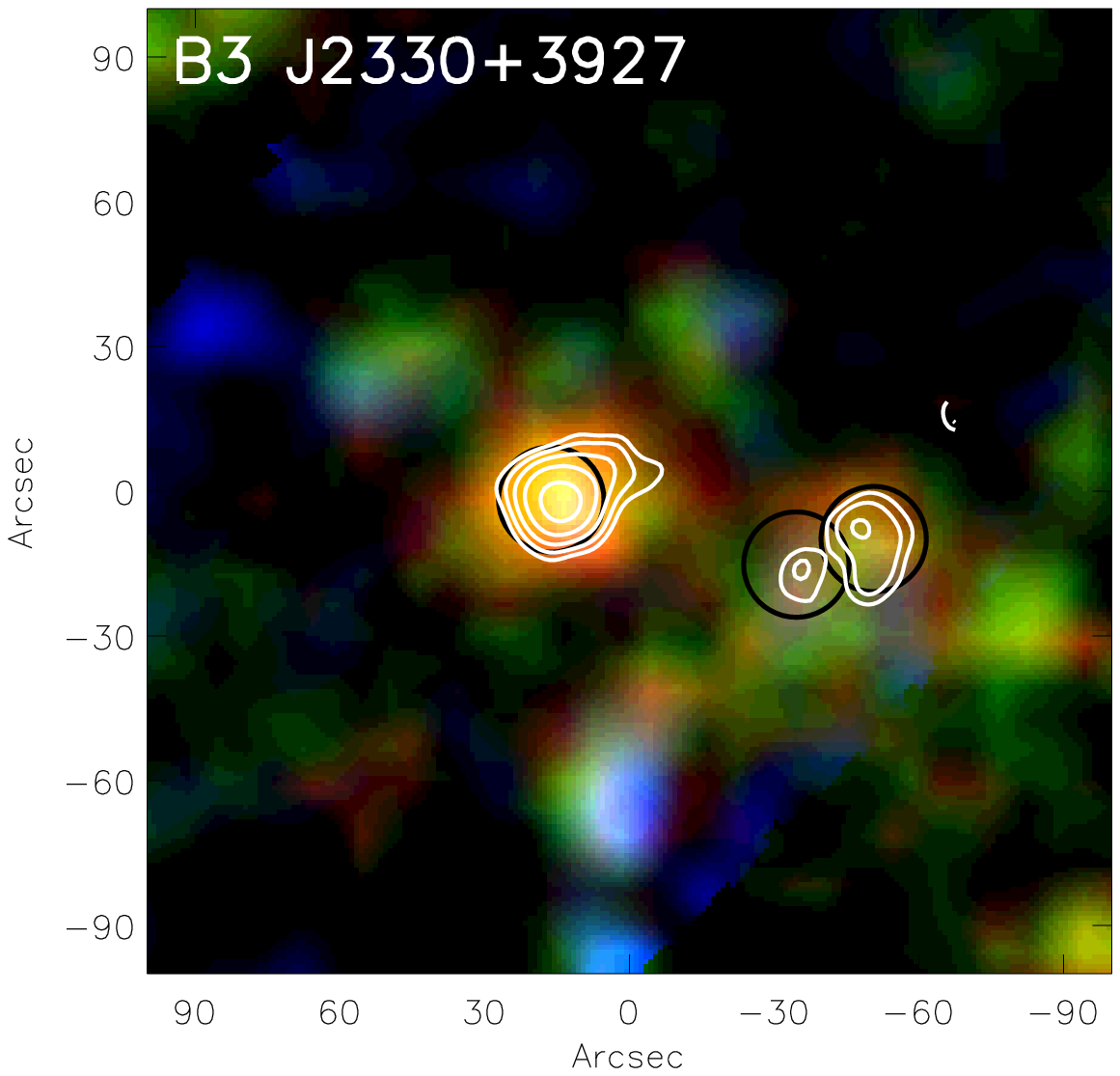,width=3.5in,angle=0}
\psfig{file=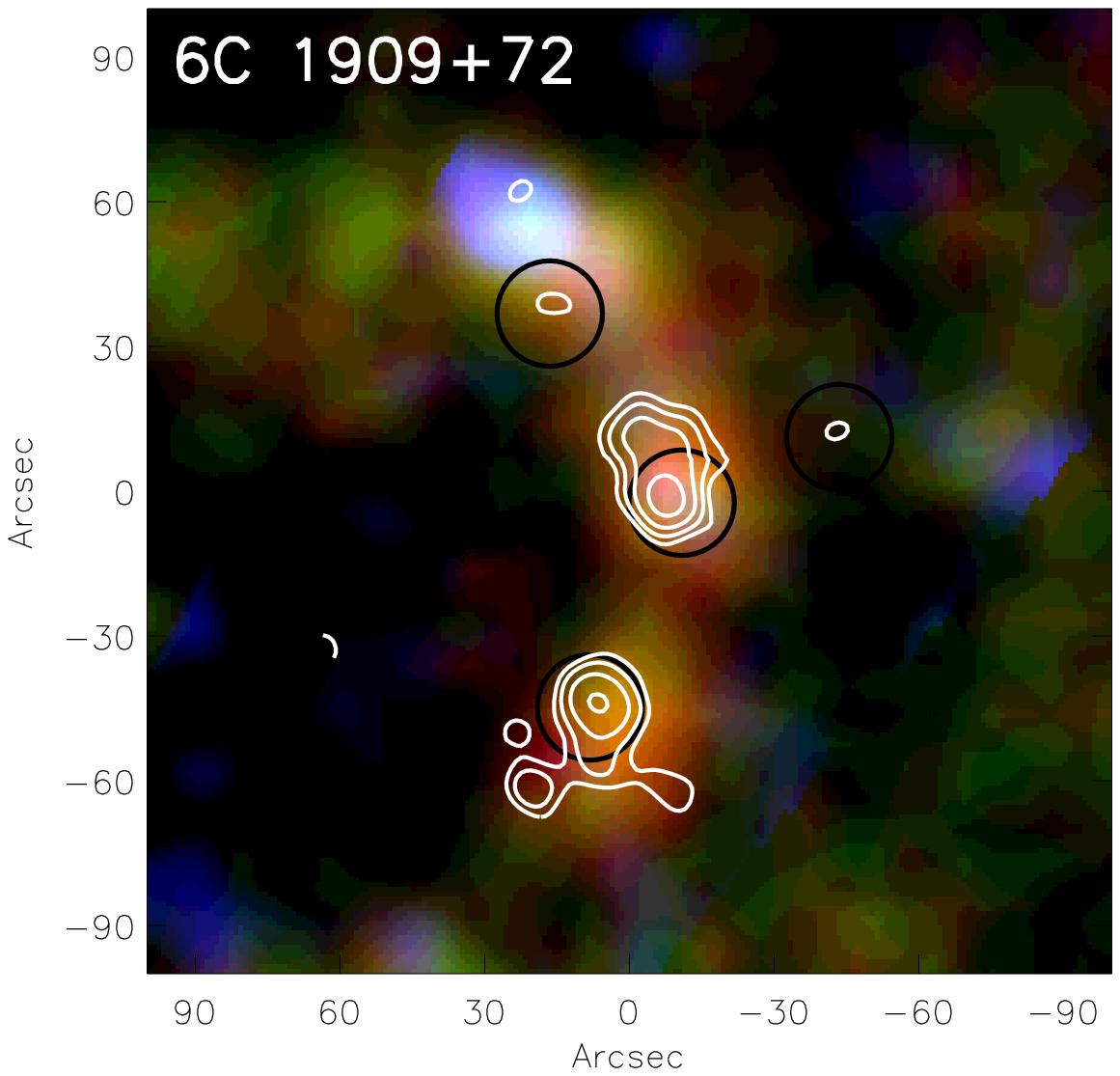,width=3.44in,angle=0}}
\vspace{-5mm} 
\noindent{\small\addtolength{\baselineskip}{-3pt}} 
\caption{850-$\mu$m continuum emission \citep[$\sqrt2$-spaced
  contours, starting at 3\,$\sigma$,][]{stevens03} for the fields
  surrounding B3\,J2330+3927 ({\it left}) and 6C\,1909+72 ({\it
    right}), superimposed on false-colour images made using the {\it
    Herschel} 160-, 250- and 350-$\mu$m imaging, described in
  \S\ref{herschelobs}, convolved to the 350-$\mu$m spatial
  resolution. N is up; E is left. The SMGs identified by
  \citet{stevens03} are circled. Unsurprisingly, these are revealed
  here as the reddest objects in each field. The red structure
  apparent in the 6C\,1909+72 field is co-aligned with both the radio
  jets (recall Fig.~\ref{fig:1909}) and a prominent extension seen
  previously at 850\,$\mu$m. A less prominent extension is seen to the
  WNW of B3\,J2330+3927.}
\label{fig:continuum}
\end{figure*} 

The black crosses represent 6C\,1909+72 and the very bright, nearby
SMG, SMM\,J190827.5+721928. Their rest-frame FIR colours are
remarkably similar in the lower two colour-colour plots. The two
galaxies differ in $S_{\rm 160}/S_{\rm 250}$, i.e.\ the colours on the
Wien side of their SEDs. The limit for the bright companion is
consistent with the known redshift of the radio galaxy, but the colour
of the radio galaxy is bluer than we might expect. On balance, the
evidence is consistent with the proposition that these two dusty
starbursts share the same node or filament\footnote{A typical filament
  at $z\sim 3.5$ spans of order $\approx 30\,h^{-1}$\,Mpc
  \citep[e.g.][]{springel05a}. SMM\,J190827.5+721928 is separated by
  45\,arcsec (335\,kpc) in the plane of the sky from 6C\,1909+72. It
  would need to lie within $\delta z\ls 0.04$ to inhabit the same
  sheet or filament.} of the cosmic web.

With the $\Delta v > 10,000$\,\kms\ of velocity coverage available to
us via the WIDAR correlator at JVLA, we should be sensitive to
\COtw\ \jonezero\ emission from bright SMGs in our target fields,
should they have a similar ratio of $S_{\rm CO}/S_{\rm 850\mu m}$ and
if -- as seems likely, at least for 6C\,1909+72 and
SMM\,J190827.5+721928 -- they lie in the same cosmic structure as the
central HzRG.

We do not detect \COtw\ \jonezero\ emission from SMM\,J233019.1+392703
in the B3\,J2330+3927 field. However, this SMG is less than half as
bright as the nearby HzRG its emission is attenuated severely (82 per
cent) by the JVLA's primary beam at 28.2\,GHz, so this does not rule
out the possibility that it shares the same structure at $z\sim
3.09$. We set a 3-$\sigma$ upper limit of $I_{\rm
  CO1-0}<0.41$\,Jy\,\kms\ where we have adopted a line width of
800\,\kms, typical for SMGs \citep[e.g.][]{greve05}.

SMM\,J190827.5+721928, in the 6C\,1909+72 field, is sufficiently
bright that we would expect to detect its \COtw\ \jonezero\ emission,
given our $\Delta v \sim 20,000$\,\kms\ [$3.39 < z < 3.75$] of velocity
[redshift] coverage in this field, if it were to have a similar ratio
of $S_{\rm CO1-0}/S_{\rm 850\mu m}$ and be part of the same structure
as the nearby HzRG. We searched the \COtw\ \jonezero\ velocity cube at
the position of the SMG, at spectral resolutions ranging from
95--380\,\kms. Despite being amongst the brightest known SMGs, its
position is ill-defined since it lacks a convincing counterpart in the
available {\it Spitzer} 3.6--24-$\mu$m imaging. The most convincing
peak lies at $\nu_{\rm obs}=25.124\pm 0.006$\,GHz, which would
correspond to $I_{\rm CO1-0} = 0.110\pm 0.033$\,Jy\,\kms\ at
$z=3.588\pm 0.001$ (after primary beam correction). The best-fit line
width ($430\pm 120$\,\kms) would be narrower than the majority of SMG
\COtw\ \jthreetwo\ lines \citep{greve05, bothwell12} and we do not
view this as a robust detection, but it would be an obvious place to
begin searching for \COtw\ \jfourthree\ with IRAM's new WideX
correlator. Several other 2--3-$\sigma$ peaks are also present, but
none with a line width even remotely commensurate with an SMG
\citep[e.g.][]{greve05}. The 1-$\sigma$ $I_{\rm CO}$ level for a line
of width 800\,\kms\ is $0.06$\,Jy\,\kms.  We are forced to conclude
that the presence of two unusually bright SMGs in this field may be
due to the chance alignment of distant starbursts.
 
The lack of a convincing detection of \COtw\ towards the brightest SMG
in the vicinity of 6C\,1909+72 casts doubt on the idea that it
co-habits a proto-cluster environment with the radio galaxy
\citep{stevens03}. However, our {\it Herschel} imaging does hint at
the presence of a dust-rich strand of the cosmic web centred on
6C\,1909+72. Fig.~\ref{fig:continuum} reveals a large
($\approx$500\,kpc) red structure, the majority of which is co-aligned
with a prominent extension ($\sim$150\,kpc, somewhat larger than the
$\sim$50-kpc scale of the northern radio jet -- recall
Fig.~\ref{fig:1909}) seen previously at 850\,$\mu$m
\citep{stevens03}. We might imagine this as a filament of gas that is
being accreted slowly onto the AGN host galaxy, but in our view it is
no less likely that this structure is due to the outflow of metal-rich
material from the radio galaxy host, heated by a network of unresolved
LIRGs. A dusty outflow could be driven by radiation pressure from a
strong nuclear starburst and/or AGN on dust
\citep*[e.g.][]{prochaska09, lehnert11, fgq12, roth12, wagner12} with
material transported well beyond the AGN host galaxy, as witnessed in
the near-IR on smaller scales for radio galaxies at $z\sim 2-3$
\citep{nesvadba08, harrison12}, for M\,87 in the Virgo cluster
\citep{simionescu08, simionescu09, werner10}, for the brightest
cluster galaxy (BCG) in the Hydra~A cluster \citep{kirkpatrick09,
  gitti11} and for BCGs more generally \citep{kirkpatrick11,
  osullivan11}. We note that the presence of $\sim$3,000-\kms\
absorption features\footnote{For a $\sim$3,000-\kms\ outflow, with a
  dynamical timescale of a few $\times 10^8$ yr, a single epoch of
  starburst and/or AGN activity would be sufficient to generate a
  structure on the scale seen here.}  against an unpolarised optical
continuum led \citet{dey97} to classify 6C\,1909+72 as a
broad-absorption-line radio galaxy, and is consistent with a powerful
outflow. Of course, we must also bear in mind the well-rehearsed
argument that radio galaxies are most easily discovered when their
jets are encountering a working surface \citep[e.g.][]{barthel96}.

%
%
\begin{figure*}
\centerline{\psfig{file=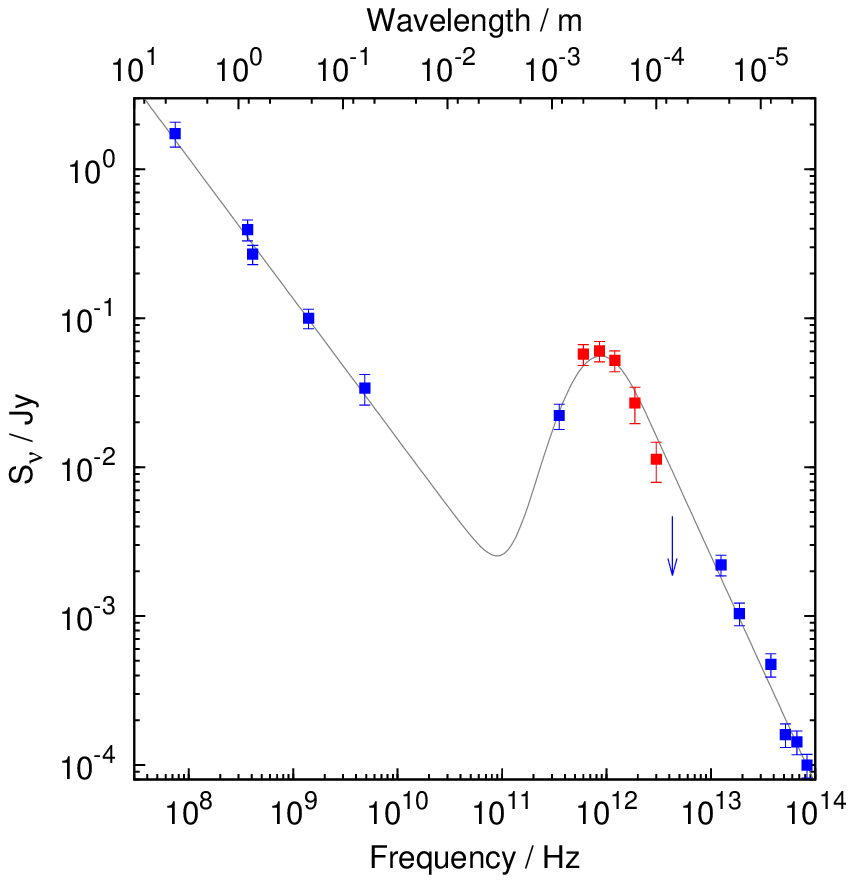,width=2.7in,angle=0} \hspace*{5mm}
\psfig{file=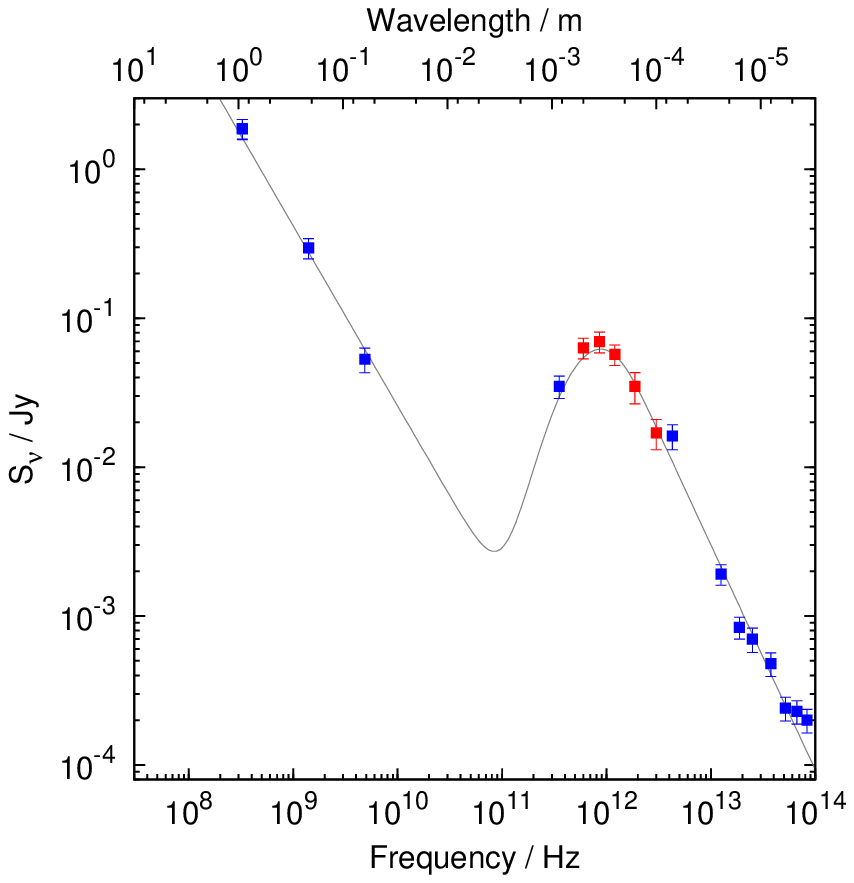,width=2.7in,angle=0}}
\vspace{-5mm} 
\noindent{\small\addtolength{\baselineskip}{-3pt}} 
\caption{Radio-through-optical SEDs of the radio galaxies,
  B3\,J2330+3927 {\it (left)} and 6C\,1909+72 {\it (right)}. Note that
  all quantities are observed rather than rest-frame. Their entire
  3\,$\mu$m to 74\,MHz SEDs can be fitted adequately (see
  \S\ref{super}) using only thermal dust and synchrotron emission
  components, whose characteristics are listed in
  Table~\ref{tab:seds}. Our new measurements from PACS and SPIRE --
  which probe the SED peaks, constraining \lir\ and \Td\ -- are shown
  in red. Lacking the spatial resolution to account for contamination
  of flux densities for B3\,J2330+3927 by component c
  (JVLA\,233024.69+392708.6), we fit to the summed flux densities of
  both galaxies.}
\label{fig:seds}
\end{figure*} 

\section{Discussion}
\label{discussion}

\subsection{On the HzRG SEDs and \Tb\ ratios}
\label{super}

Fig.~\ref{fig:seds} shows the observed radio-through-optical SEDs of
6C\,1909+72 and B3\,J2330+3927. For both galaxies, the photometry
between 3\,$\mu$m and 74\,MHz can be described adequately with a model
comprising only synchrotron and thermal dust emission, following
\citet{kovacs10}. Table~\ref{tab:seds} lists the results of a
simultaneous fit to the following free parameters: the dominant cold
dust temperature, \Td, and the power-law index, $\gamma$, of a dust
temperature distribution, $dM_{\rm d}/dT_{\rm d} \propto T_{\rm
  d}^{-\gamma}$, designed to offer a physically motivated treatment of
the Wien side of the thermal emission spectrum (in
\citeauthor{kovacs10}, $T_{\rm c}\equiv T_{\rm d}$, the
low-temperature cutoff of the distribution), the dust mass, \Md\ (for
a characteristic photon cross-section to mass ratio, $\kappa_{\rm
  850\mu m}=0.15$\,m$^2$\,kg$^{-1}$, from \citealt*{dunne03}, where
the frequency dependence of the dust emissivity, $\beta$, was fixed to
+1.5) and the synchrotron power-law index, $\alpha$. The resulting
measurements of \lir\ and \qir\ \citep[as defined by][but where
$S_{\rm IR}$ is measured across $\lambda_{\rm rest}=8-1,000\,\mu$m, as
is \lir]{helou85} are also listed.

Both radio galaxies have warmer dust temperatures, $T_{\rm d}^{\rm
  HzRG}\sim 45$\,{\sc k}, than those seen for similarly luminous,
dusty starbursts at $z\sim 3$ \citep[$T_{\rm d}^{\rm SMG}\sim
35$\,{\sc k} when calculated using a power-law temperature
distribution, as we have here --][]{magnelli12}. Determining the
contribution of the AGN to \lir\ is a long-standing problem that we
cannot hope to solve without dramatically improved spatial
resolution. The relatively warm dust does imply significant AGN
contributions to \lir\ for these two radio galaxies, though even this
cannot be asserted with certainty since the brightest 100--160-$\mu$m
emission in the B3\,J2330+3927 system (\S\ref{b3}) is centred near the
gas-rich companion (which may contain its own AGN, if it is anything
like the interacting system seen towards 4C\,60.07 --
\citealt{ivison08}). On the other hand, the detection of dust and
molecular gas towards the radio galaxy hosts makes it likely that
significant levels of star formation are taking place in both. Given
the strong dependence of \lir\ on \Td\ \citep[e.g.][]{eales00}, we
might expect that the fraction of \lir\ due to the AGN, $L_{\rm
  IR}^{\rm AGN}/L_{\rm IR}\approx 1- \eta (T_{\rm d}^{\rm HzRG}/T_{\rm
  d}^{\rm SMG})^{-6}$, where $\eta\approx 0.5$ is the fractional AGN
contribution to \lir\ estimated for typical SMGs
\citep[e.g.][]{frayer98}.  This suggests a $\approx$90-per-cent
AGN-related contribution to \lir\ for our HzRGs, which still leaves
room for vigorous starburst activity in the radio galaxy hosts, with
$\rm SFR \sim 500\,M_{\odot}\,yr^{-1}$ \citep{kennicutt98a}.

The super-thermal \Tb\ ratio, $r_{4-3/1-0} > 2.7 (3\sigma)$, observed
towards the AGN host galaxy, B3\,J2330+3927, reveals the presence of
highly excited molecular gas, given the excitation requirements of
\COtw\ \jfourthree\ ($E_{\rm 4-3} / k \sim 55$\,{\sc k} and $n_{\rm
  crit} \sim 1.9\times 10^4$\,cm$^{-3}$). Even for 6C\,1909+72, the
\Tb\ ratio, $r_{4-3/1-0}=0.76\pm 0.18$, is significantly higher than
the average value seen for SMGs by \citet{bothwell12}. IRAM PdBI and
VLA observations of 4C\,41.17 at $z=3.80$ revealed a similar story,
with $r_{4-3/1-0}$ likely super-thermal \citep{papadopoulos05}.  In
comparison, the same line ratio averaged over the entire Milky Way is
$r_{\rm 4-3/1-0}\sim 0.1-0.2$ \citep*{fixsen99}, indicative of
quiescent gas in the interstellar medium (ISM).

%
%
\begin{table}
  \centering
  \caption{\Tb\ ratios and parameters from
    spectral energy distributions.\label{tab:seds}}
  \begin{tabular}{lccc}
    \hline \hline 
Parameter & 6C\,1909+72 & B3\,J2330+3927\\
\hline

$r_{4-3/1-0}$&$0.76\pm 0.18$&$3\sigma>2.7$\\
\hline
$\chi_{\rm red}^2$&1.14&1.00\\
\Td\ ({\sc k})&$45.7\pm 1.3$&$41.4\pm 1.1$\\ 
log \Md\  (M$_{\odot}$)&$9.37\pm 0.04$&$9.26\pm 0.17$\\
$\gamma$&$5.51\pm 0.05$&$5.54\pm 0.04$\\
$\alpha$&$-1.22\pm 0.04$&$-0.95\pm 0.07$\\
\qir&$-1.02\pm0.04$&$-0.40\pm 0.11$\\ 
log \lir\ (L$_{\odot}$)&$13.69\pm 0.08$&$13.52\pm 0.07$\\
    \hline
  \end{tabular}
\end{table}
 
In contrast to B3\,J2330+3927, the \Tb\ ratio that we find for its
companion galaxy, JVLA\,233024.69+392708.6 ($r_{4-3/1-0} = 0.43\pm
0.11$) is consistent with the average value seen for SMGs and with
star-forming regions within nearby starburst galaxies such as M\,82
and NGC\,253, which have values in the range $\sim$0.5--0.8
\citep{gusten03, mao00}. The \Tb\ ratio in JVLA\,233024.69+392708.6 is
thus consistent with gas heated by UV radiation from star formation,
i.e.\ less extreme ISM conditions than those of the radio galaxy.

\Tb\ ratios as high as that observed towards B3\,J2330+3927 are
expected in environments where the gas energetics are dominated by
X-rays. These can penetrate large columns of gas and maintain a high
kinetic gas temperature, $T_{\rm k}$, much deeper into molecular
clouds than far-ultraviolet (FUV) photons \citep*[e.g.][]{meijerink06,
  schleicher10, vdw10}, with this gas thermally decoupled from the
cooler dust.  Powerful radio-loud AGN are often found to have high
X-ray luminosities, capable of powering extended X-ray-dominated
regions (XDRs); while B3\,J2330+3927 has no X-ray measurements, this
is one possible explanation for the observed $r_{4-3/1-0}$.

Another possible explanation is shock-excited gas
\citep[e.g.][]{flower10} caused by its radio jet ramming into the
molecular environment, bearing in mind that the mechanical power of a
radio jet can be orders of magnitude larger than that of its
synchrotron emission. An example of this is observed in the powerful
radio galaxy, 3C\,293, which has Milky Way-like levels of star
formation and a very low X-ray luminosity \citep[see also][regarding
the central region of M\,51]{matsushita04}. Its super-thermal \COtw\
(as measured in \COtw\ \jonezero\ and \jfourthree) has been attributed
to turbulent heating from the dissipation of shocks caused by the
interaction of its powerful radio jet with its molecular gas
\citep{papadopoulos08, nesvadba10, guillard12} -- an interaction that
seemingly also drives a $\sim$1,400-\kms\ outflow in H\,{\sc i}
\citep{morganti03}. Such mechanisms are often proposed to provide the
feedback needed to reconcile models of galaxy formation with a variety
of observations \citep[e.g.][and references therein]{granato04,mn12,
  fabian12}.

A large fraction of the radio emission from B3\,J2330+3927 comes from
its core ($\sim$50 per cent at 8.4\,GHz) and it displays an unusually
one-sided jet: `one of the most asymmetric radio structures ever
reported for a type {\sc ii} AGN' \citep{perez-torres05}. This may be
due to relativistic Doppler beaming \citep[e.g.][]{rw90} or because of
the launch mechanism \citep*[e.g.][]{cbt96} but if a survey of such
galaxies were to find relatively high \Tb\ ratios relative to galaxies
hosting two-sided jets then it would suggest a more prosaic reason: a
blockage that causes an unusually high fraction of the mechanical
energy from AGN-driven outflow to be deposited into the molecular gas.

According to \citet{wb11}, the kinetic energy and momentum of a jet
can be transfered to dense gas with a high efficiency (10--70 per
cent), causing turbulent heating and shock excitation of the molecular
gas and -- in extreme cases -- driving gas outflows and inhibiting
star formation. The kinetic energy of the radio jet in B3\,J2330+3927
can be estimated following \citet{punsly05} and \citet{punsly08},
assuming $S_{\rm 151MHz} = 0.90$\,Jy \citep[from $S_{\rm 365MHz} =
0.394$\,Jy and $\alpha^{\rm 365MHz}_{\rm 74MHz} = -0.93$][though see
also \citealt{birzan08} and \citealt{cavagnolo10}]{douglas96,
  cohen07}. The jet's kinetic luminosity, $Q$, is then $\sim 2 \times
10^{46}$\,erg\,s$^{-1}$, roughly two orders of magnitude larger than
the \COtw\ \jfourthree\ luminosity, $L'_{\rm CO4-3} = 3.4 \times
10^{44}$\,erg\,s$^{-1}$. This simplistic approach assumes that the
151-MHz emission comprises optically thin radiation from the lobes,
whereas a substantial fraction may come from a Doppler-boosted
jet. Nevertheless, over the $\sim$10-Myr lifetime of the radio source
we can expect a few $10^{60}$\,erg to be deposited into the
multi-phase ISM \citep[enough to quench a cooling flow for
$\sim10^9$\,yr, as argued by][]{mcnamara05}. The turbulent kinetic
energy of the molecular gas can be calculated through $E^{\rm
  turb}_{\rm kin} = \frac{3}{2} M_{\rm H_2} {\sigma_{\rm H_2}}^{2}$,
where $\sigma_{\rm H_2} =$ {\sc fwhm}$/2\sqrt{\rm 2\,ln\,2}$ is the
velocity dispersion of the gas. From \S\ref{b3}, $M_{\rm H_2} <
2.3\times 10^{10}$\,M$_{\odot}$ so if $\sigma_{\rm CO1-0} \sim
\sigma_{\rm CO4-3} = 350$\,\kms\ and is solely due to the velocity
dispersion of the gas (i.e.\ no components due to rotation or bulk
outflow), then $E^{\rm turb}_{\rm kin} < 10^{59}$\,erg. We can thus
conclude, as did \citep[as did][for an analagous
situation]{nesvadba08} that the radio jet carries sufficient energy to
explain the observed CO characteristics. The jet can influence the gas
properties in precisely the way required to stop star formation,
though questions clearly remain about the ultimate fate of the gas and
deeper observations are required to search for any high-velocity
molecular component \citep[see, e.g.,][]{polletta11}.

Taking another approach, if we adopt a fiducial jet advance speed,
$v_{\rm jet} \sim 0.1c$, then the gas is heated to $T_{\rm
  shock}\approx 3/16\,\mu\,m_{\rm H}/k\,v_{\rm jet}^2 \approx
10^{10}$\,{\sc k}, where $\mu$ is the reduced mass and $k$ is
Boltzmann's constant. For a large fraction of the gas that has been
shock-heated by the jet, the density will be low
\citep[$10^{-2}$--$10^{-3}$\,cm$^{-3}$,][]{wb11}.  This
over-pressurised gas will expand \citep{begelman89}, running into the
denser, ambient ISM, which will be engulfed and destroyed by the
resulting shock \citep*[e.g.][]{klein94}.  The velocity of the shock
that is driven into the dense clouds, $v_{\rm s,cloud}=v_{\rm wind}
(\rho_{\rm wind}/\rho_{\rm cloud})^{1/2}$, where $\rho_{\rm wind}$ is
the density of the hot plasma created by the outward-moving radio jet
and $\rho_{\rm cloud}$ is the density of the engulfed ISM. For typical
molecular clouds, $\rho_{\rm cloud}\sim$10$^4$\,cm$^{3}$, so we arrive
at $v_{\rm s,cloud}\sim 20$--40\,\kms\ \citep{begelman89}.  Such
molecular shocks would be significant emitters in high-$J$ CO lines
\citep{flower10}.

At what rate must the gas must be shock-heated to maintain this CO
emission? For a C-type shock wave with $v_{\rm s,cloud}\sim 40$\,\kms\
(assuming a lower velocity or a denser ambient ISM will make little
difference, and a J-type shock will produce a line luminosity almost
two orders of magnitude fainter), we find that to explain $L'_{\rm
  CO4-3}$ requires a mass shock rate of
$\sim30$\,M$_{\sun}$\,yr$^{-1}$ \citep{flower10}. Over the age of the
jet, only a few $\times 10^9$\,M$_{\sun}$ of gas needs to be shock
heated, perhaps much less if this is a special, short-lived phase due
to the initial passage of the radio jet. The gas may return to a
higher density following the passage of the shock \citep{cooper08},
perhaps re-forming H$_2$ \citep*{guillard09}, so the amount of
shock-heated gas required to explain the observed \Tb\ ratios is not
restrictive.

Whatever the excitation mechanism, the \COtw\ optical depth must be
moderate to low ($\tau_{1-0} \ls 1$) since in the optically thick
regime the \Tb\ of a transition is equal to the excitation
temperature, i.e.\ $T_{\rm b} \sim T_{\rm ex}$ (with the latter being
the same for all levels when thermalised -- see
\citealt{papadopoulos12a}). In the optically thin case, $T_{\rm
  b}^{4-3}/T_{\rm b}^{1-0} \propto \tau_{4-3}/\tau_{1-0}$ and since
$\tau_{\rm J+1, J} \propto (J+1)$ (at least up to $J\sim 7$), we find
$T_{\rm b}^{1-0} < T_{\rm b}^{4-3}$ (true for thermalised gas, but
also for sub-thermal gas if the effect due to optical depth is
sufficiently large). Optically thin \COtw\ emission can be achieved if
the medium is highly turbulent, as is the case for the Galactic Centre
-- a local region where the \COtw\ ladder is observed to be
super-thermal.  The super-thermal \Tb\ ratio observed towards
B3\,J2330+3927 is thus likely an indication of mechanical energy
deposited into the ISM from the radio jet and/or the ongoing
starburst, inducing a degree of turbulence sufficient to make the
\COtw\ lines optically thin, allied with a high density ($\ge n_{\rm
  crit}$) and a relatively high kinetic temperature; moreover, this
must happen globally, rather than in some small fraction of the gas,
as seen in the nucleus of M\,82, for example \citep{knapp80, weiss01}.

\subsection{The merger fraction amongst starbursting HzRGs}
\label{mergers}

The degree to which various populations of star-forming galaxies are
driven by violent mergers of gas-rich galaxies or by the accretion of
cold gas streams -- sometimes smooth, sometimes clumpy -- via dark
matter filaments into spiral disks \citep*{dekel09b} has been a topic
of intense debate in recent years. Even for SMGs -- the most extreme
star-formation events in the Universe, where the evidence for
merger-driven activity was thought by many to be overwhelming
\citep{engel10} -- the seeds of doubt have been sown.

What of radio galaxies? We cannot generalise to the entire population,
even though that population is small in number: the time-consuming
interferometric observations of molecular gas that are capable of
laying bare the processes leading to star formation have generally
been obtained towards systems from which rest-frame FIR emission has
already been detected, i.e.\ we are limited to systems pre-selected to
be undergoing intense starbursts. Even so, it is interesting to add
our findings to those studies of $z\gs 2$ radio galaxies in the
literature and explore the overall statistics. Taking the eight
systems listed in \S\ref{intro} in order of decreasing redshift:

\begin{itemize}
\item The interferometric \COtw\ \jfourthree\ study of 4C\,41.17, at
  $z=3.80$, by \citet{debreuck05} revealed activity driven by a merger
  between two massive, gas-rich galaxies separated by $\sim$2\,arcsec
  ($\sim$13\,kpc) and $\sim$400\,\kms.
\item High-resolution interferometric imaging of 4C\,60.07 at $z=3.79$
  by \citet{ivison08} revealed two components of roughly equal
  integrated flux, separated by $\sim$30\,kpc -- evidence of an
  early-stage merger between the host galaxy of an actively-fueled
  black hole (the HzRG), and a gas-rich starburst/AGN, caught prior to
  its eventual equilibrium state. This separation is typical of SMGs
  with multiple radio identifications \citep{ivison07} and for the
  intense burst of star formation near first passage in merger
  simulations \citep*[e.g.][]{springel05b}.
\item \citet{smith10} use rest-frame UV spectroscopy to show that
  6C\,1909+72, at $z=3.53$, comprises two galaxies separated by
  $\sim$1,300\,\kms\ along the line of sight, with the radio-loud AGN
  hosted by the more distant component.
\item TN\,J0121+1320 at $z=3.52$ was detected in \COtw\ \jfourthree\
  by \citet{debreuck03ar}. There were no obvious signs of a gas-rich
  companion, but the available sensitivity, velocity coverage and
  spatial resolution were poor compared to recent studies.
\item We have shown here that the immediate environment of
  B3\,J2330+3927 at $z=3.09$ contains at least two gas-rich
  components, again separated by $\sim$30\,kpc.
\item \citet{nesvadba09} presented interferometric \COtw\ imaging of
  the $z=2.58$ radio galaxy, TXS\,0828+193, finding two large gas
  reservoirs separated by 720\,km\,s$^{-1}$ along a shared line of
  sight roughly 80\,kpc SW of the AGN host galaxy, which was itself
  detected only in synchrotron at 96.6\,GHz.
\item 53W002 at $z=2.39$ was observed and detected in \COtw\
  \citep{yamada95, scoville97} without knowledge of its FIR
  luminosity, though it was known to sit amongst a clump of
  Ly-$\alpha$ emitting galaxies \citep{pascarelle96} through which the
  radio galaxy is suspected to be assembling a large fraction of its
  eventual stellar mass through mergers \citep[e.g.][]{motohara01}. At
  least one of these -- 330\,kpc and $\gs$300\,\kms\ from 53W002 -- is
  FIR-luminous \citep{smail03hzrg}. Based on their \COtw\ imaging,
  \citet{scoville97} concluded `it is clear that the emission is
  resolved', but this was not corroborated by
  \citet{alloin00}. However, with a $\sim$6-arcsec synthesised beam
  the latter data would be insensitive to close companions.
\item Finally, {\it Hubble Space Telescope} near-IR imaging of
  MRC\,0152$-$209 at $z=1.92$ revealed a morphology consistent with an
  advanced-stage merger \citep{pentericci01} where we would expect the
  gas to have been driven into a compact configuration, well within
  the 7--10-arcsec {\sc fwhm} synthesised beam of the available \COtw\
  observations \citep{emonts11}.
\end{itemize}

Overall, we have six mergers or close galaxy pairs, one ambiguous
case, and one case where no merger activity can be discerned (albeit
with relatively poor sensitivity and resolution available). The
statistics suggest, then, that violent interactions are as ubiquitous
amongst starbursting high-redshift radio galaxies as they are amongst
SMGs \citep{engel10}.

\section{Conclusions}
\label{conclusions}

Combining our new interferometric \COtw\ data for two distant radio
galaxies with those for six more, reported elsewhere, we find that
activity in starbursting radio-loud AGN at high redshift is driven by
the merger or interaction of two or more systems in which significant
masses of molecular gas and stars have already formed, rather than the
steady\footnote{Often denoted `secular', which in astronomy indicates
  that a process occurs continuously, as opposed to a discrete or
  periodic event such as a merger.} accretion of cold gas from the
cosmic web.

We introduce a new colour-colour diagnostic plot to constrain the
redshifts of several distant, dusty galaxies found previously in our
target fields. We conclude that the SMG south of 6C\,1909+72 likely
shares the same node or filament, but we fail to detect this
FIR-luminous galaxy in \COtw, despite $\sim$20,000\,\kms\ of velocity
coverage.

We introduce a new colour-colour diagnostic plot, exploiting the power
of 350--850-$\mu$m photometry to constrain the redshifts of several
distant, dusty galaxies found previously in our target fields. We
conclude that the bright SMG near 6C\,1909+72 likely shares the same
node or filament of the cosmic web at $z\sim 3.5$ as the signpost
AGN. However, we fail to detect this SMG in \COtw\ \jonezero, despite
our $\sim$20,000\,\kms\ of velocity coverage.

Also in the 6C\,1909+72 field, we find an unusually large, red dust
feature, aligned with the radio jet. We suggest that metal-rich
material may have been dispersed on $\gs$100-kpc scales by a
collimated outflow, reminiscent of the jet-oriented metal enrichment
seen in X-ray observations of local cluster environments.

We find that \COtw\ brightness temperature ratios in the host galaxies
of radio-loud AGN are significantly higher than those seen in
similarly intense starbursts where AGN activity is less pronounced. In
our most extreme example, the super-thermal \Tb\ ratio suggests that
significant energy is being deposited rapidly into the molecular gas
via X-rays and/or mechanical (`quasar-mode') feedback from the AGN,
leading to a high degree of turbulence {\it globally} and a low
optical depth in \COtw\ -- feedback that will lead to the cessation of
star formation on a timescale commensurate with that of the jet
activity, $\ls$10\,Myr.

\section*{Acknowledgements}

We are grateful to Arjun Dey, Axel Weiss and Padelis Papadopoulos for
sharing their wisdom regarding super-thermal \Tb\ ratios and
unpublished optical spectroscopy. We appreciate the remarkable efforts
of the NRAO staff that have significantly upgraded what many of us
already regarded as the finest telescope ever built. In particular, we
thank Frazer Owen and Gustaaf Van Moorsel for their invaluable help
with the data used here. IRS acknowledges support from STFC and
through a Leverhulme Senior Fellowship. NS is the recipient of an
Australian Research Council Future Fellowship. TRG acknowledges
support from STFC, as well as IDA and DARK. This work is based on
observations carried out with the Karl Janksy Very Large Array. The
NRAO is a facility of the NSF operated under cooperative agreement by
Associated Universities, Inc. It is also based on observations carried
out with the IRAM Plateau de Bure Interferometer. IRAM is supported by
INSU/CNRS (France), MPG (Germany) and IGN (Spain).

\bibliographystyle{mnras}
\bibliography{ivison}

\bsp

\end{document}